# NEW CYCLOTRON RESONANCES

*V.A. Buts[1,2,3], V.V. Kuzmin[1], A.P. Tolstoluzhsky[1]*
[1]*National Science Center "Kharkov Institute of Physics and Technology", Kharkiv, Ukraine;*
[2]*Institute of Radio Astronomy of NAS of Ukraine, Kharkiv, Ukraine;*
[3]*V.N. Karazin Kharkiv National University, Kharkiv, Ukraine*
*E-mail: vbuts@kipt.kharkov.ua*

The possibilities and conditions of effective interaction, in particular acceleration, of charged particles by the field of plane electromagnetic wave in the presence of an external constant magnetic field are considered. It is shown that the well-known conditions of cyclotron resonances require generalization. New conditions for the resonant interaction of charged particles are formulated, which contain not only the strength of the external magnetic field (as the well-known conditions of cyclotron resonances) but also the field strength of the wave. strengths. It is shown that new resonance conditions open up new possibilities for effective particle acceleration.

## INTRODUCTION

Acceleration of charged particles in a vacuum seems to be a tempting prospect. There are a large number of works (both theoretical and experimental) devoted to this problem (see, for example, [1 - 8]). They also indicate the advantages of such acceleration and the problems that one has to face when solving such tasks.

In the presence of a constant magnetic field, the situation changes qualitatively. Cyclotron resonances appear ($\omega = \vec{k}\vec{v} + \omega_H / \gamma$). When using them, an effective interaction of waves and particles is possible. Particularly attractive is the auto-resonance acceleration scheme. However, to realization this scheme when using laser radiation fields, abnormally large external magnetic fields are required. It should be noted that only external magnetic field intensity ($\omega_H$) is included in cyclotron resonance conditions. There is no wave field strength under these conditions. This is due to the fact that the theory of cyclotron resonances developed when almost always the wave strength parameter ($\varepsilon = eE / mc\omega$) was small. Therefore, it was not necessary to take it into account.

The wave intensity appeared only in the study of nonlinear cyclotron resonances. With the development of laser technology, the situation could change. As indicated above, the use of cyclotron resonances seemed simply impossible. In addition to lasers, sources of intense electromagnetic radiation appeared, such as, for example, CRM. However, only the usual conditions of cyclotron resonances were still used (see above).

It is clear that when the wave power parameter becomes significant, the usual conditions for cyclotron resonance must be modified. In this conditions, both the strength of the external magnetic field and the strength of the fields with which the particles interact must be present. This is especially true for the case of laser fields, when the cyclotron frequency is much lower than the frequency of laser radiation ($\omega_H/\omega \ll 1$). This work is devoted to the analysis of the use of both the usual conditions of cyclotron resonance and new modified conditions.

## 1. STATEMENT OF THE PROBLEM AND BASIC EQUATIONS

Consider a charged particle that moves in an external constant magnetic field $H_0$ directed along the axis $z$ and in the field of a plane electromagnetic wave, which in the general case has the following components:

$$\mathcal{E} = \mathrm{Re}(\mathbf{E}\exp(i\omega t - i\mathbf{kr})), \quad \mathbf{H} = \mathrm{Re}\left(\frac{c}{\omega}[\mathbf{kE}]\exp(i\omega t - i\mathbf{kr})\right), \tag{1}$$

where $\mathbf{E} = E_0\boldsymbol{\alpha}$, $\boldsymbol{\alpha} = \{\alpha_x, i\alpha_y, \alpha_z\}$ is wave polarization vector.

Without limiting of generality, we can assume that the wave vector $\mathbf{k}$ has only two nonzero components $k_x$ and $k_z$. In dimensionless variables $\mathbf{p} \to \mathbf{p}/mc$, $\tau \to \omega t$, $\mathbf{r} \to \frac{\omega}{c}\mathbf{r}$, particle equations of motion can be reduced to:

$$\frac{d\mathbf{p}}{d\tau} = \left(1 - \frac{\mathbf{kp}}{\gamma}\right)\mathrm{Re}\left(\boldsymbol{\mathcal{E}}e^{i\psi}\right) + \frac{\omega_H}{\gamma}[\mathbf{ph}] + \frac{\mathbf{k}}{\gamma}\mathrm{Re}\left[(\mathbf{p}\boldsymbol{\mathcal{E}})e^{i\psi}\right], \tag{2}$$

$$\mathbf{v} = \frac{d\mathbf{r}}{d\tau} = \frac{\mathbf{p}}{\gamma}, \quad \dot{\psi} = \frac{d\psi}{d\tau} = 1 - \frac{\mathbf{kp}}{\gamma},$$

where $\mathbf{h} = \mathbf{H}/H_0$, $\omega_H = eH/mc\omega$, $\boldsymbol{\mathcal{E}} = \varepsilon_0\boldsymbol{\alpha}$, $\varepsilon_0 = (eE_0/mc\omega)$, $\psi = \tau - \mathbf{kr}$, $\mathbf{k}$ is unit vector in the direction of wave propagation, $\gamma = (1+\mathbf{p}^2)^{1/2}$ is particle energy, $\mathbf{p}$ is its momentum.

Multiplying the first of equations (2) by $\mathbf{p}$, we obtain the following equation describing the change of particle energy:

$$\frac{d\gamma}{d\tau} = \mathrm{Re}\left(\mathbf{v}\boldsymbol{\mathcal{E}}e^{i\psi}\right), \tag{3}$$

Using equations (3), from the system of equation (2) we find the integral of motion:

$$\mathbf{p} - \mathrm{Re}\left(i\boldsymbol{\mathcal{E}}e^{i\psi}\right) + \omega_H[\mathbf{rh}] - \mathbf{k}\gamma = \mathrm{const}. \tag{4}$$

## 2. ANOTHER FORM OF MAIN EQUATIONS

We firstly consider the case of wave propagation along an external magnetic field $H_0$. Then the vector equation (2) and equation (3) can be conveniently rewritten in the following form:

$$\dot{p}_x = \dot{\psi}\varepsilon_x\cos\psi + \omega_H(p_y/\gamma), \quad \dot{p}_y = -\dot{\psi}\varepsilon_y\sin\psi - \omega_H(p_x/\gamma),$$

$$\dot{\gamma} = \frac{1}{\gamma}\left(p_x\varepsilon_x\cos\psi - p_y\varepsilon_y\sin\psi\right), \tag{5}$$

where $\varepsilon_x = \alpha_x\varepsilon_0$, $\varepsilon_y = \alpha_y\varepsilon_0$.

Note that the value $\gamma\dot\psi = C$ is an integral. Then the equations for the transverse components of the particle pulse can be issued separately in closed form:

$$p'_x = \varepsilon_x \cos\psi + \Omega p_y, \quad p'_y = -\varepsilon_y \sin\psi - \Omega p_x, \tag{6}$$

here $p' = \dfrac{dp}{d\psi}$, $\Omega = (\omega_H / \gamma\dot\psi)$.

The solutions of the system of equations (6) under the condition $\Omega = const$ can be presented in the analytical form:

$$p_x = A\sin\Omega\psi + B\cos\Omega\psi + \frac{\varepsilon_1}{(1-\Omega^2)}\sin\psi,$$

$$p_y = C\sin\Omega\psi + D\cos\Omega\psi + \frac{\varepsilon_2}{(1-\Omega^2)}\cos\psi \tag{7}$$

where:

$$A = p_{x0}\sin\Omega\psi_0 + p_{y0}\cos\Omega\psi_0 + \sin\Omega\psi_0\sin\psi_0\frac{\varepsilon_1}{\Omega^2-1} + \cos\Omega\psi_0\cos\psi_0\frac{\varepsilon_2}{\Omega^2-1},$$

$$B = p_{x0}\cos\Omega\psi_0 - p_{y0}\sin\Omega\psi_0 + \cos\Omega\psi_0\sin\psi_0\frac{\varepsilon_1}{\Omega^2-1} - \sin\Omega\psi_0\cos\psi_0\frac{\varepsilon_2}{\Omega^2-1},$$

$$C = -B, \quad D = A, \quad \varepsilon_1 = \varepsilon_x + \Omega\varepsilon_y, \quad \varepsilon_2 = \varepsilon_y + \Omega\varepsilon_x.$$

Using (5), it is easy to find analytical expressions for the longitudinal momentum:

$$p_z = \frac{1}{2\gamma\dot\psi}\left[p_x^2 + p_y^2 - (p_{x0}^2 + p_{y0}^2)\right] + p_{z0}. \tag{9}$$

Similarly, we define the expression for the particle energy $\gamma$, for example, for a wave with linear polarization:

$$\gamma = \frac{1}{2\gamma\dot\psi}\left[p_x^2 + p_y^2 - (p_{x0}^2 + p_{y0}^2)\right] + \gamma_0. \tag{11}$$

It can be seen from the expressions obtained for the components of the particle momentum that both the magnitude of the momentum and the particle energy are periodic functions of the phase. Therefore, the effective transfer of energy from laser radiation to charged particles will occur only in a limited space (or for a limited time). The value of this space (or time interval) can be found by determining, for example, the dependence of the phase on time. It is easy to do. So, for linear polarization, it is easy to find the following expression for the phase

$$\psi \sim \left(6C\gamma^2/\varepsilon 0^2\right)^{1/3}\tau^{1/3}. \tag{12}$$

From this formula it is seen that the dimensionless time of effective particle acceleration is proportional to the integral $(\gamma - p_z) = C$. It will be the greater, the greater this integral.

The above expressions for the components of momentum and for energy were obtained under the conditions when $\Omega \neq 1$, i.e. when there are no autoresonance conditions. If the autoresonance conditions are satisfied $\Omega = 1$, then using the system of equations (6) we obtain the following expressions for the momenta in the case of a wave with circular polarization:

$$p_x = \varepsilon_0 (\psi - \psi_0) \cos \psi, \quad p_y = -\varepsilon_0 (\psi - \psi_0) \sin \psi. \tag{13}$$

## 3. DYNAMICS OF PARTICLES AT CYCLOTRON RESONANCES

Above, expressions have been obtained for cases when it is possible to obtain solutions in an explicit analytical form. In these expressions, cyclotron resonances are not revealed explicitly. Below we obtain a general system of equations in which cyclotron resonances can be explicitly single out.

For this, it is convenient to introduce new variables $p_\perp, p_z, \theta, \xi$ and $\eta$, which will explicitly display the dynamics of particles in a constant magnetic field:

$$p_x = p_\perp \cos\theta,\, p_y = p_\perp \sin\theta,\, p_z = p_\parallel,\, p_\perp = \sqrt{p_x^2 + p_y^2},\, x = \xi - \frac{p_\perp}{\omega_H}\sin\theta,\, y = \eta + \frac{p_\perp}{\omega_H}\cos\theta. \tag{14}$$

We substitute these variables in the vector equation (2). Expanding the right-hand side of the obtained equations in a series of Bessel functions, we have:

$$\frac{dp_\perp}{d\tau} = \varepsilon_0 \left[ \begin{array}{l} \left((\alpha_x - \alpha_y)(1 - k_z v_z) + \alpha_z k_x v_z\right) \times \sum_{n=-\infty}^{+\infty} \left(\frac{n}{\mu} J_n(\mu) \cos(\theta_n)\right) + \\ + \alpha_y (1 - k_z v_z) \sum_{n=-\infty}^{+\infty} (J_{n+1}(\mu)) \cos(\theta_n) \end{array} \right], \tag{15}$$

where $\mu = k_x p_\perp / \omega_H$, $\theta_n = k_z z + k_x \xi - \tau - n\theta$, $J_n = J_n(\mu)$, $J'_n = dJ_n(\mu)/d\mu$.

Similarly, we obtain expressions for $\dot{p}_z,\, \dot{\gamma},\, \dot{\theta},\, \dot{\xi},\, \dot{\eta},\, \dot{\theta}_n$:

$$\frac{dp_z}{d\tau} = \varepsilon_0 \alpha_z \sum_{n=-\infty}^{\infty} \left(1 - \frac{n\omega_H}{\gamma}\right) J_n \cos\theta_n + \varepsilon_0 k_z v_\perp \sum_{n=-\infty}^{\infty} \left(\alpha_x \frac{n}{\mu} J_n - \alpha_y J'_n\right) \cos\theta_n, \tag{16}$$

$$\frac{d\gamma}{d\tau} = \varepsilon_0 \sum_{n=-\infty}^{\infty} \left(\alpha_x v_\perp \frac{n}{\mu} J_n - \alpha_y v_\perp J'_n + \alpha_z v_z J_n\right) \cos\theta_n, \tag{17}$$

$$\frac{d\theta}{d\tau} = \frac{\varepsilon_0 \left(1 - \vec{k}\vec{v}\right)}{p_\perp} \sum_{n=-\infty}^{+\infty} \left(\alpha_x J'_n(\mu) - \alpha_y \frac{n}{\mu} J_n(\mu)\right) \sin(\theta_n) + \\ + \frac{k_x \varepsilon_0}{p_\perp} \sum_{n=-\infty}^{+\infty} J'_n(\mu)\left(\alpha_y v_y \cos(\theta_n) + (\alpha_x v_x + \alpha_z v_z) \sin(\theta_n)\right) - \frac{\omega_H}{\gamma}, \tag{18}$$

$$\frac{d\xi}{dt} = -\frac{\varepsilon_o}{\omega_H} \sum_{n=-\infty}^{\infty} \left(\alpha_y (1 - k_z v_\parallel) + \alpha_y k_x v_\perp \frac{n}{\mu}\right) J_n \sin\theta_n, \tag{19}$$

$$\frac{d\eta}{dt} = -\frac{\varepsilon_o}{\omega_H} \sum_{n=-\infty}^{\infty} \left(\alpha_x (1 - k_z v_\parallel) J_n + k_x (\alpha_y v_\perp J'_n - \alpha_z v_\parallel J_n)\right) \cos\theta_n, \tag{20}$$

$$\frac{dx}{dt} = \frac{p_\perp \cos(\theta)}{\gamma}, \quad \frac{dy}{dt} = \frac{p_\perp \sin(\theta)}{\gamma}, \quad \frac{dz}{dt} = \frac{p_z}{\gamma}. \tag{21}$$

Further on the right hand we leave only the resonance terms, i.e. terms for which the parameters satisfy the condition of one of the well-known cyclotron resonances: $\omega = \vec{k}\vec{v} + n\omega_H$. Using one of these conditions it is possible to obtain equations describing the motion of a particle under conditions of isolated resonance:

$$\dot{p}_\perp = \frac{1}{p_\perp}(1 - k_z v_z) W_n \cdot \varepsilon_0 \cos\theta_n, \quad \dot{p}_z = \frac{\varepsilon_0}{\gamma} k_z W_n \cos\theta_n, \quad \dot{\gamma} = \frac{\varepsilon_0}{\gamma} W_n \cdot \cos\theta_n$$

$$\dot{\theta}_n \equiv \Delta_n \equiv k_z v_z + n\frac{\omega_H}{\gamma} - 1, \tag{22}$$

where: $W_n \equiv \alpha_x p_\perp \frac{n}{\mu} J_n - \alpha_y p_\perp J_n' + \alpha_z p_z J_n$.

Carrying out the expansion $\Delta_n(\gamma)$ near the resonance value $\gamma_0$ from the last equations of system (22) we obtain a closed system for describing the dynamics of particles in the isolated resonance:

$$\dot{\theta}_n = \frac{k_z^2 - 1}{\gamma_0} \delta\gamma, \quad \delta\dot{\gamma} = \frac{\varepsilon_0}{\gamma_0} W_n \cos\theta_n. \tag{23}$$

Using these equations, it is easy to find the magnitude of particle energy gain in isolated cyclotron resonance:

$$\delta\gamma = 4\sqrt{\varepsilon_0 W_n / (1 - k_z^2)}. \tag{24}$$

## 4. NEW CYCLOTRON RESONANCES

The system of equations (14) - (20) was studied in sufficient detail in [5, 6, 8]. This system is convenient for analysis when the parameter $\varepsilon$ is small. In this case, the averaging method was used to analyze this system. However, system (14) - (20) is strictly valid for any parameter value.

The form of equations (14)-(21), which describes the dynamics of particles, is convenient for finding out resonant conditions. An effective energy exchange between wave and particle will occur when one or more of the terms in the right-hand sides of equations (14) - (21) change slowly. It can be seen that this will happen when the condition is met:

$$\dot{\theta}_n = k_z v_z + k_x \dot{\xi} - 1 - n\dot{\theta} = 0 \tag{25}$$

Further we will be mainly interested in the dynamics of particles in laser fields. It means that in real conditions the dimensionless cyclotron frequency will also be a small parameter ($\omega_H \ll 1$). In addition, in most cases, we will be interested in the dynamics of relativistic particles ($\gamma \gg 1$). Note that condition (25) takes into account the dynamics of the leading center, which substantially depends on the electric field strength of the laser radiation. In the special case ($\dot{\xi}=0$), conditions (25) contain the well-known conditions of cyclotron resonance. We consider some particular new resonance conditions:

1. The simplest case is when the parameters of the fields and particles satisfy the following relations

$$n=0 \quad k_z=0,\ k_x=1 \quad \omega_H \ll 1 \quad \varepsilon_x=\varepsilon_z=0. \tag{26}$$

Then condition (25) can be represented as:

$$\frac{\varepsilon_y}{\sqrt{\omega_H p_\perp}}\cos\left(\frac{p_\perp}{\omega_H}-\frac{\pi}{4}\right)\sin\theta_0 = -\sqrt{\frac{\pi}{2}}. \tag{27}$$

It can be seen that the resonance condition substantially depends on the wave strength parameter ($\varepsilon_y$).

2. If parameters of fields and particles satisfy relation:

$$n=0;\ k_z \to 1,\ k_x \ll 1;\ \omega_H \ll 1;\ \varepsilon_x=\varepsilon_z=0,$$

then the expression for cyclotron resonance takes the form:

$$v_z + \frac{k_x \varepsilon_y}{\omega_H \gamma^2}\sin\theta_0 = 1. \tag{28}$$

Using resonance conditions (25), as well as equations from system (15) - (21), we can obtain the following equation for describing the phase dynamics in the vicinity of resonance:

$$\ddot{\theta}_0 + \frac{\varepsilon_y k_x v_\perp^2}{2\gamma}\cos\theta_0 = 0. \tag{29}$$

Equation (29) is the equation of a mathematical pendulum. Analysis of such equations and consequence of a similar analysis can be found in [5, 6, 9].

3. The most interesting case is when the parameters of the wave and particles satisfy the conditions:

$$n=\mu \gg 1;\ k_z \to 1,\ k_x \sim \left(1/\gamma^2\right) \ll 1;\ \omega_H \ll 1;\ \varepsilon_x=\varepsilon_z=0. \tag{30}$$

The importance of this case is due to the fact that it allows us to analyze the resonance at large values of number ($n \gg 1$). Besides, this case corresponds to the situation when the number of the Bessel function is equal to the argument of the Bessel function. In this case, as is known, the Bessel function decreases most slowly with the growth of its number and argument ($J_n(n) \sim 1/\sqrt[3]{n}$). The resonance condition for this case has the form:

$$\dot{\theta}_n = k_x\dot{\xi}-n\dot{\theta} = \frac{n\omega_H^2 - 2\varepsilon_y k_x^2 p_\perp J_n \sin\theta_n}{\gamma \omega_H} \equiv \Delta, \tag{31}$$

$$\Delta(\gamma_0)=0.$$

Here $\gamma_0$ is value of energy at which the exact resonance condition is satisfied ($\Delta(\gamma_0) = 0$). To describe the dynamics of the phase, we can derive the equation:

$$\ddot{\theta}_n + \Omega^2 \cos\theta_n \sin\theta_n - \Omega_1^2 \cos\theta_n = 0, \qquad (32)$$

where: $\Omega^2 = \dfrac{2v_\perp^2 \varepsilon_y^2 J_n^2}{\gamma \omega_H}$; $\Omega_1^2 = \dfrac{n\omega_H v_\perp \varepsilon_y J_n}{\gamma^2}$.

Equation (31) is also the equation of a nonlinear pendulum and has the integral:

$$\frac{\dot{\theta}_n^2}{2} + \frac{\Omega^2}{2}\sin^2\theta_n - \Omega_1^2 \sin\theta_n = C = const. \qquad (33)$$

Analysis of this integral shows that the maximum phase velocity can be estimated by $\dot{\theta}_{max} \approx \Omega$. Using this estimation, it is easy to determine the value of addition to the particle energy that they obtain when interacting with the wave under resonance conditions:

$$\dot{\theta}_n = \Delta(\gamma_0) + \left(\frac{\partial \Delta}{\partial \gamma}\right)_{\gamma_0} \delta\gamma,$$

$$(\delta\gamma)_{max} = (\dot{\theta}_{max})/(\partial\Delta/\partial\gamma) \approx \varepsilon_y J_n \sqrt{\omega_H \gamma^3}. \qquad (34)$$

Comparing this additive with those obtained under conditions of known cyclotron resonances (24), we can see that it can be more significant.

The lines of new resonances in the parameter plane ($\varepsilon, \omega_H$) are represented in Fig.1 and Fig2

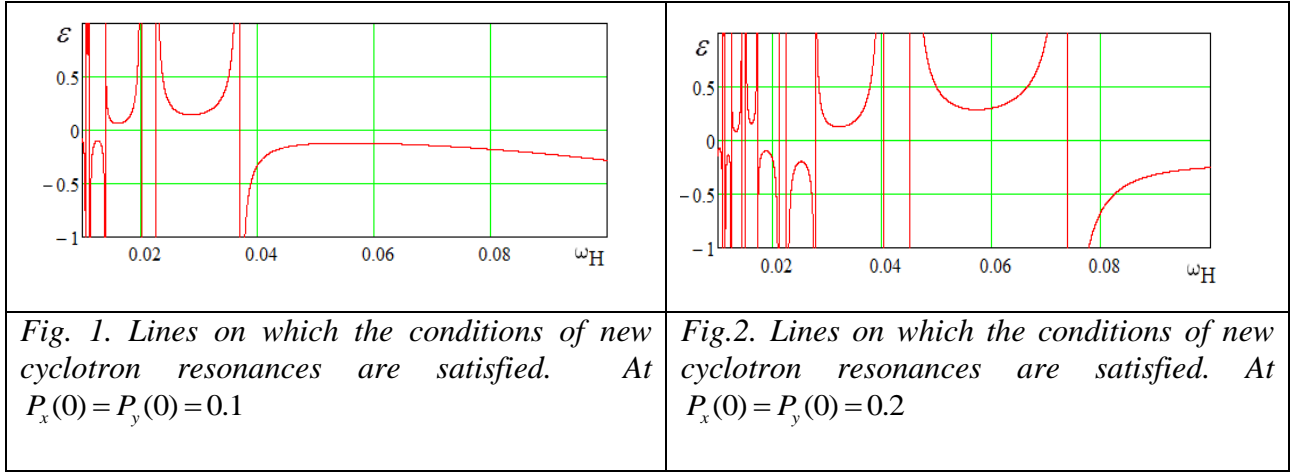

Fig. 1. Lines on which the conditions of new cyclotron resonances are satisfied. At $P_x(0) = P_y(0) = 0.1$

Fig.2. Lines on which the conditions of new cyclotron resonances are satisfied. At $P_x(0) = P_y(0) = 0.2$

## 5. NUMERICAL ANALYSIS

For a wave propagating along the direction of the external magnetic field, analytical solutions of equations (5.6) are found for momentum and coordinates of particle in an implicit form as a function of phase $\psi$.

Besides, the integral $\gamma\dot\psi = C$ is break down when the wave propagates at an angle to the external magnetic field $(k_\perp = k_x \neq 0)$. Therefore, a numerical analysis of equations (2) was carried out to investigate the dynamics of charged particles in the field of the plane electromagnetic wave and in the external constant magnetic field $H_0$ directed along the axis $z$. The cases of linear and circular polarization of the wave field are considered. Since we are mainly interested in particle acceleration, we consider this process at sufficiently large initial values of the longitudinal momentum of the particles and small values of the transverse momentum (for small values of the transverse momentum, the parameter $\mu \ll 1$).

The analysis was carried out at the initial values of the longitudinal momentum $p_{z0} = 10$; the transverse momenta were chosen equal to $p_{x0} = p_{y0} = 0.1$. The initial values of the transverse coordinates are selected in accordance with the values of the transverse momenta and the external constant magnetic field, the initial coordinate $z_0 = z(t=0) = 0$. The accuracy of the calculations was controlled using the integral (4). In all the numerical studies, the value of the integral was preserved with a sufficient degree of accuracy: the value of deviation from the integral did not exceed the values $10^{-7} - 10^{-6}$ for the coordinates and momenta of charged particles of the order $10^3$.

As follows from the above formulas, the value of the longitudinal momentum $p_z \gg p_\perp$ therefore the value $p_z$ practically coincides with the energy value $\gamma$.

If the initial values of the momenta of the charged particles are such that the condition is satisfied $C = \gamma - p_z$, where $C = \gamma\dot\psi = const$ is the integral of particle motion, a scheme of autoresonant interaction of particles with laser fields at $\omega_H = \gamma\dot\psi$ can be realized.

Figs. 3, 4 shows graphs of the dependence of the longitudinal and transverse pulses, as well as the longitudinal and transverse coordinates of the particles on time under conditions of autoresonance for a wave with circular polarization $\varepsilon_x = \varepsilon_y = \varepsilon_0$, $\varepsilon_z = 0$ for the field amplitude $\varepsilon_0 = 0.75$ and $\omega_H = \gamma_0\dot\psi_0 = 0.5087$.

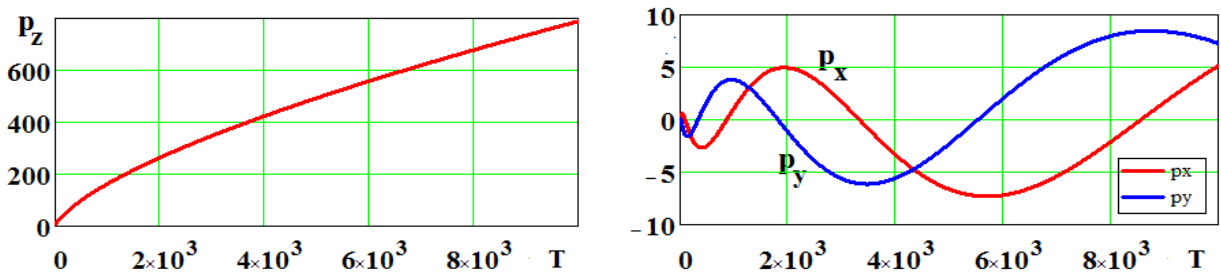

*Fig. 3. Dependences of the longitudinal $p_z$ and transverse momenta on time $T = \tau/2\pi$. Circular polarization*

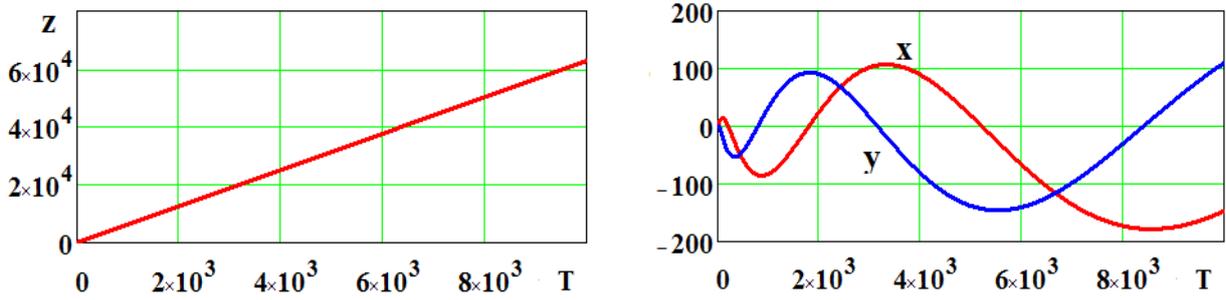

*Fig.4. Dependences of the longitudinal $z$ and transverse coordinates $x$, $y$ on time $T = \tau/2\pi$. Circular polarization*

In the case of linear polarization $\varepsilon_x = \varepsilon_z = 0$, $\varepsilon_y = \varepsilon_0$, the graphs of the dependence of the longitudinal and transverse pulses, as well as the longitudinal and transverse coordinates of the particles on time under conditions of autoresonance are shown in Figs. 5, 6.

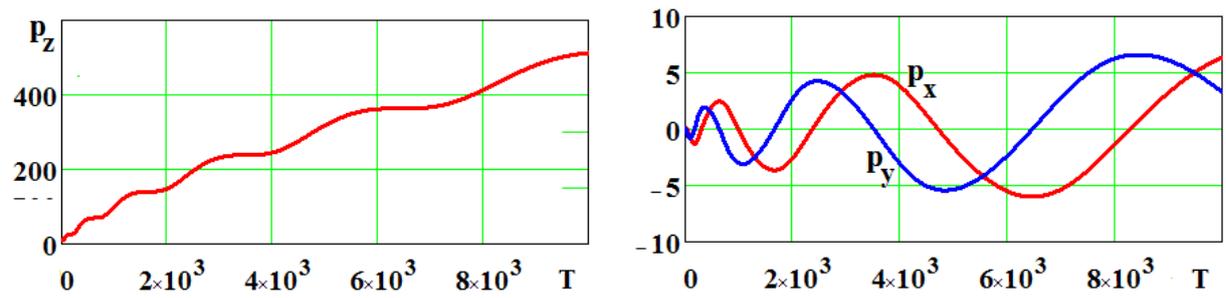

*Fig. 5. Dependences of the longitudinal $p_z$ and transverse momenta on time $T = \tau/2\pi$. Linear polarization.*

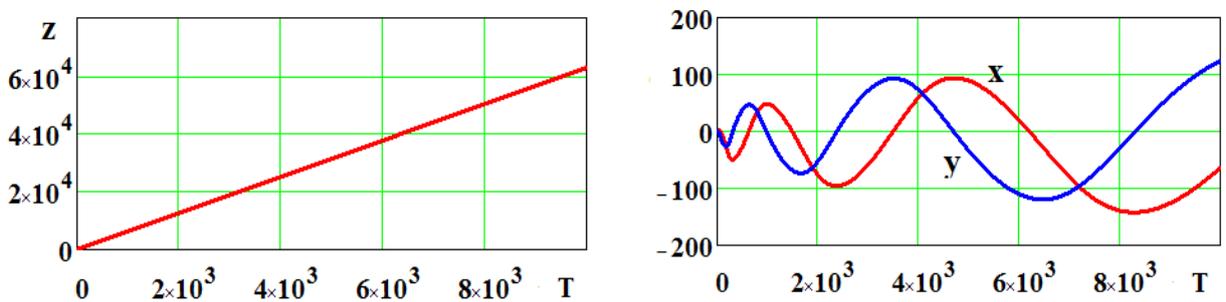

*Fig.6. Dependences of the longitudinal $z$ and transverse coordinates $x$, $y$ on time $T = \tau/2\pi$. Linear polarization.*

As can be seen from these graphs, the maximum values of the longitudinal and transverse momenta with circular polarization are approximately two times higher than their values with linear polarization. The dependence of the longitudinal coordinate on time, as expected ($\beta_z \approx c$), has not practically changed. The oscillation period of the transverse coordinates and momenta is approximately two times large than in the case of linear polarization.

In the case of oblique propagation ($k_x = 0.075$) of the linearly polarized wave $\varepsilon_y = \varepsilon_0$ $\varepsilon_x = \varepsilon_z = 0$, the dependences of the longitudinal and transverse coordinates and momenta of the particle for cyclotron frequency $\omega_H = \gamma_0 \dot{\psi}_0$ and parameter values are shown in Figs. 7, 8.

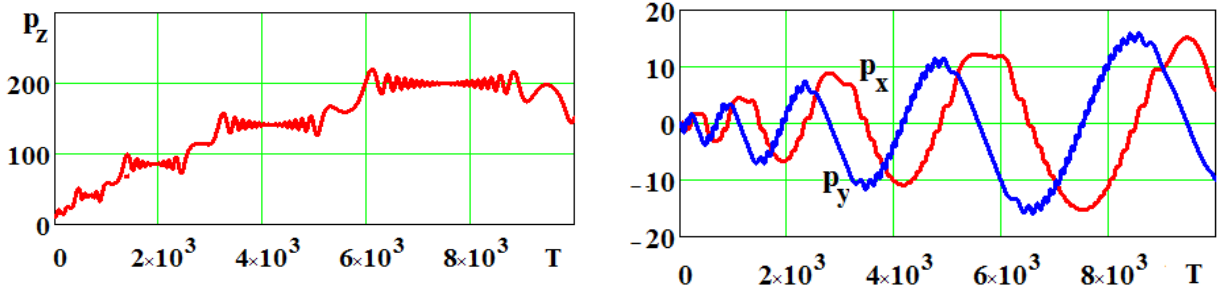

*Fig. 7. Ddependences of the longitudinal $p_z$ and transverse momenta on time $T = \tau/2\pi$. Linear polarization*

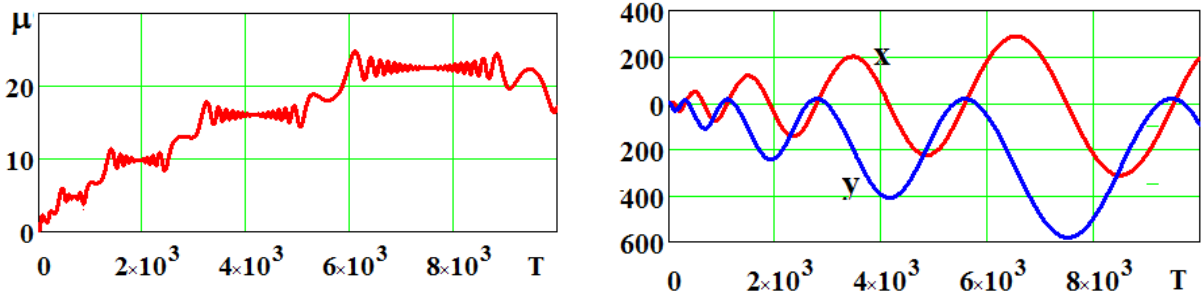

*Fig.8. Dependences of the parameter $\mu$ and transverse coordinates $x$, $y$ on time $T = \tau/2\pi$. Linear polarization*

From the graphs in Figs. 7, 8 it is seen that the interaction of a charged particle with a field is resonant character. The time intervals at which the parameter $\mu$ oscillates around a certain average value are clearly distinguished. In accordance with the change in the parameter $\mu$, the average energy of the oscillations of the energy of the charged particle also changes. At the same time, contribution to the energy increment gives not only one harmonic with a fixed number $n$, but also adjacent harmonics $n-1$, $n+1$:

$$\Delta\gamma(n) \approx \varepsilon_y \sum_{k=n-1}^{n+1} J_k(\mu)\sqrt{\omega_H \gamma^3}. \tag{34}$$

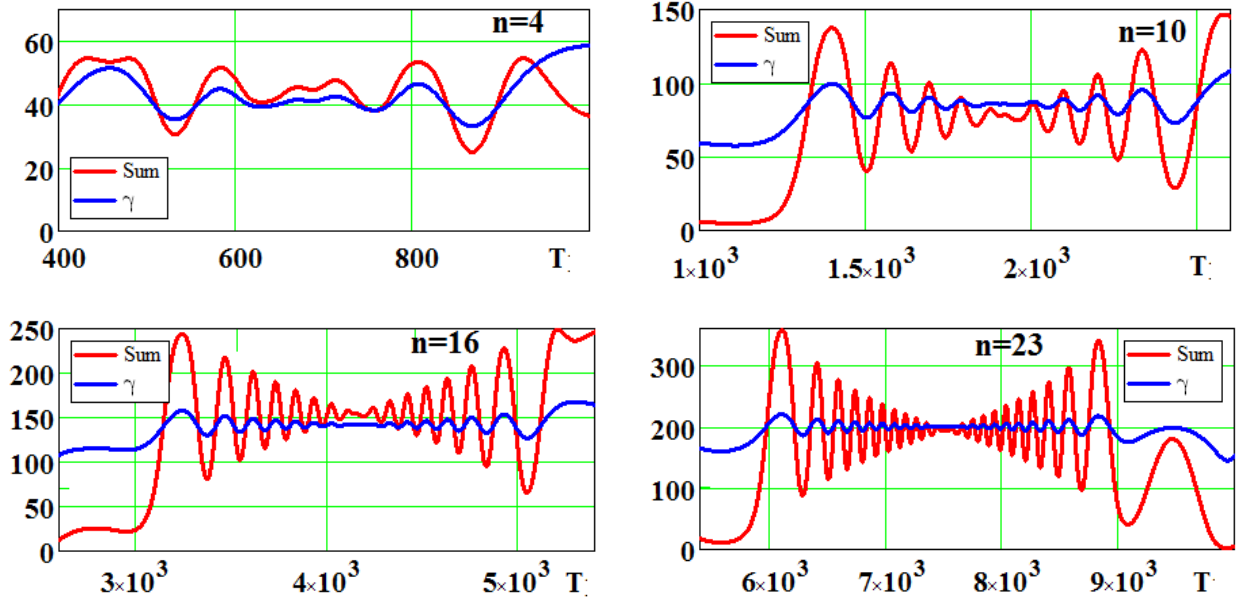

*Fig. 9. Dependences of particle energy on time $T = \tau/2\pi$ for different regions change of parameter $\mu$. Blue color indicates the curve of the particle energy versus time obtained by numerically solving the system of equations (2). Red color indicates the curve of the dependence of the particle energy on time found by the formula (34)*

As can be seen from the graphs in Fig. 9, we can speak of a sufficiently good qualitative agreement between the results of the numerical calculation of the system of equations (2) and the results of evaluation by formula (34).

| 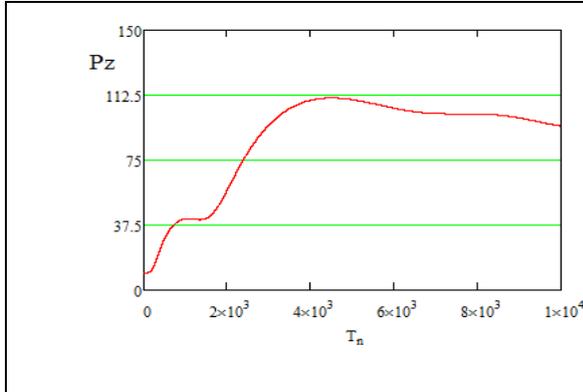 | 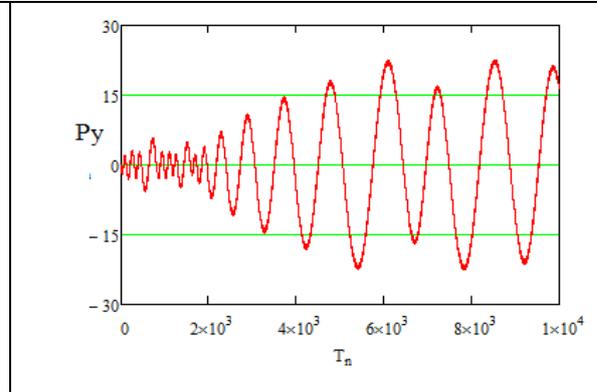 |
|---|---|
| *Fig.10. New resonances. Longitudinal momentum evolution. At $n=0$; $k_z \to 1$, $k_x \ll 1$; $\omega_H = 0.1$; $P_z(0)=10$; $P_x(0)=P_y(0)=0$; $\varepsilon_y = 1.1$; $\varepsilon_x = \varepsilon_z = 0$* | *Fig.11. New resonances. Transverse momentum evolution. At $n=0$; $k_x=1$, $k_z=0$; $\omega_H=0.1$; $P_z(0)=0$; $P_x(0)=1$; $P_y(0)=0$; $\varepsilon_y = -0.405$; $\varepsilon_x = \varepsilon_z = 0$* |

Figures 10-11 show the results of a numerical study of the dynamics of longitudinal and transverse pulses under conditions of new resonances. The conditions of the usual cyclotron resonances cannot be satisfied.

## CONCLUSIONS

Let's state the most important results of this work.

1. It is shown that the well-known conditions of cyclotron resonance should be generalized. The generalization is that these conditions include both the strength of the external magnetic fields and the field strength of the electromagnetic waves with which the particles interact. The use of these new resonant conditions makes possible to implement a scheme of resonant interaction of particles even with laser radiation fields in vacuum.

2. If the initial parameters of charged particles are such that the condition $C = \gamma - p_z = \omega_H$, where $C = \gamma \dot{\psi} = const$ is satisfied, where $C$ is the integral of particle motion, then a scheme of autoresonant interaction of particles with laser fields in vacuum can be implemented.

3. Conducted numerical studies confirm qualitatively and quantitatively the key results of analytical studies within the framework of this model.